# Quantum Cost Optimization for Reversible Sequential Circuit

Md. Selim Al Mamun
Jatiya Kabi Kazi Nazrul Islam University
Trishal, Mymensingh-2220, Bangladesh

David Menville
Pascack Valley High School
New Jersey, United States of America

*Abstract*—Reversible sequential circuits are going to be the significant memory blocks for the forthcoming computing devices for their ultra low power consumption. Therefore design of various types of latches has been considered a major objective for the researchers quite a long time. In this paper we proposed efficient design of reversible sequential circuits that are optimized in terms of quantum cost, delay and garbage outputs. For this we proposed a new 3*3 reversible gate called SAM gate and we then design efficient sequential circuits using SAM gate along with some of the basic reversible logic gates.

*Keywords—Flip-flop; Garbage Output; Reversible Logic; Quantum Cost*

I. INTRODUCTION

In recent years, reversible computing has emerged as a promising technology. The primary reason for this is the increasing demands for lower power devices. In the early 1960s R. Landauer [1] demonstrated that losing bits of information causes loss of energy. Information is lost when an input cannot be recovered from its output. In 1973 C. H. Bennett [2] showed that energy dissipation problem can be avoided if the circuits are built using reversible logic gates.

Reversible logic has the feature to generate one to one correspondence between its input and output. As a result no information is lost and there is no loss of energy [3]. Although many researchers are working in this field, little work has been done in the area of sequential reversible logic. In the current literature on the design of reversible sequential circuits, the number of reversible gates is used as a major metric of optimization [4]. The number of reversible gates is not a good metric of optimization as reversible gates are of different type and have different quantum costs [5]. In this paper, we presented new designs of reversible sequential circuits that are efficient in terms of quantum cost, delay and the number of garbage outputs.

This paper is organized as follows: Section 2 presents some basic definitions related to reversible logic. Section 3 describes some basic reversible logic gates and their quantum implementation. Section 4 introduces our proposed gate 'Selim Al Mamun' (SAM) gate. Section 5 describes the logic synthesis of sequential circuits and comparisons with other researchers. Finally this paper is concluded with the Section 6.

II. BASIC DEFINITIONS

In this section, some basic definitions related to reversible logic are presented. We formally define reversible gate, garbage output, delay in reversible circuit and quantum cost of reversible in reversible circuit.

*A. Reversible Gate*

A Reversible Gate is a k-input, k-output (denoted by k*k) circuit that produces a unique output pattern for each possible input pattern [6]. If the input vector is $Iv$ where $Iv = (I_{1,j}, I_{2,j}, I_{3,j}, ...., I_{k-1,j}, I_{k,j})$ and the output vector is $Ov$ where $Ov = (O_{1,j}, O_{2,j}, O_{3,j}, ..., O_{k-1,j}, O_{k,j})$, then according to the definition, for each particular vector $j$, $Iv \leftrightarrow Ov$.

*B. Garbage Output*

Every gate output that is not used as input to other gates or as a primary output is garbage. Unwanted or unused outputs which are needed to maintain reversibility of a reversible gate (or circuit) are known as Garbage Outputs. The garbage output of Feynman gate [7] is shown in Fig. 1 with *.

*C. Delay*

The delay of a logic circuit is the maximum number of gates in a path from any input line to any output line. The definition is based on two assumptions: (i) Each gate performs computation in one unit time and (ii) all inputs to the circuit are available before the computation begins.

In this paper, we used the logical depth as measure of the delay proposed by Mohammadi and Eshghi [8]. The delay of each 1x1 gate and 2x2 reversible gate is taken as unit delay 1. Any 3x3 reversible gate can be designed from 1x1 reversible gates and 2x2 reversible gates, such as CNOT gate, Controlled-V and Controlled-V$^+$ gates (V is a square-root-of NOT gate and V$^+$ is its hermitian). Thus, the delay of a 3x3 reversible gate can be computed by calculating its logical depth when it is designed from smaller 1x1 and 2x2 reversible gates.

*D. Quantum Cost*

The quantum cost of a reversible gate is the number of 1x1 and 2x2 reversible gates or quantum gates required in its design. The quantum costs of all reversible 1x1 and 2x2 gates are taken as unity [9]. Since every reversible gate is a combination of 1 x 1 or 2 x 2 quantum gate, therefore the quantum cost of a reversible gate can be calculated by counting the numbers of NOT, Controlled-V, Controlled-V$^+$ and CNOT gates used.

III. QUANTUM ANALYSIS OF DIFFERENT REVERSIBLE GATES

Every reversible gate can be calculated in terms of quantum cost and hence the reversible circuits can be measured in terms of quantum cost. Reducing the quantum cost from reversible circuit is always a challenging one and works are still going on in this area. This section describes some popular reversible gates and quantum equivalent diagram of each reversible gate.

*A. Feynman Gate*

Let $I_v$ and $O_v$ are input and output vector of a 2*2 Feynman gate where $I_v$ and $O_v$ are defined as follows: $I_v = (A, B)$ and $O_v = (P = A, Q = A \oplus B)$. The quantum cost of Feynman gate is 1. The block diagram and equivalent quantum representation for a 2*2 Feynman gate are shown in Fig. 1.

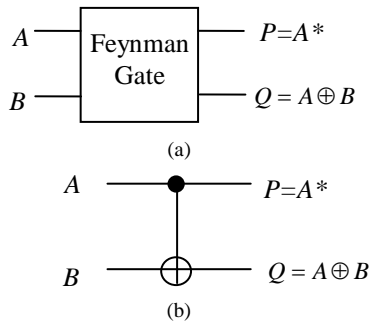

Fig. 1. (a) Block diagram of 2x2 Feynman gate and (b) Equivalent quantum representation

*B. Double Feynman Gate*

Let $I_v$ and $O_v$ are input and output vector of a 3*3 Double Feynman gate (DFG) where $I_v$ and $O_v$ are defined as follows: $I_v = (A, B, C)$ and $O_v = (P = A, Q = A \oplus B, R = A \oplus C)$. The quantum cost of Double Feynman gate is 2 [10]. The block diagram and equivalent quantum representation for 3*3 Double Feynman gate are shown in Fig. 2.

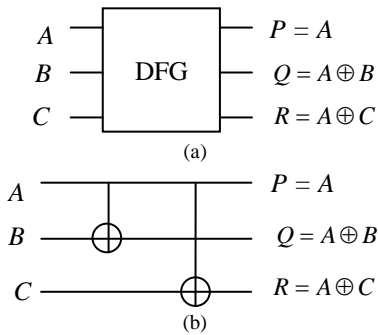

Fig. 2. (a) Block diagram of 3x3 Double Feynman gate and (b) Equivalent quantum representation.

*C. Toffoli Gate*

The input vector, $I_v$ and output vector, $O_v$ for 3*3 Toffoli gate (TG) [11] can be defined as follows: $I_v = (A, B, C)$ and $O_v = (P = A, Q = B, R = AB \oplus C)$. The quantum cost of Toffoli gate is 5.

The block diagram and equivalent quantum representation for 3*3 Toffoli gate are shown in Fig. 3.

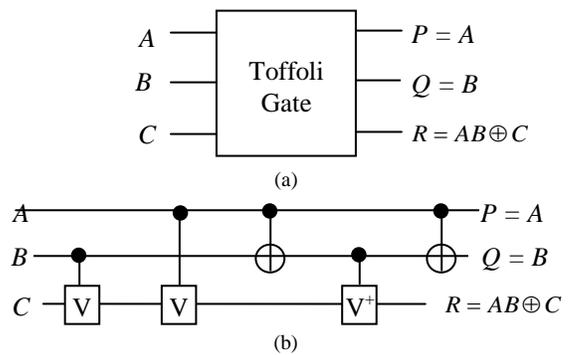

Fig. 3. (a) Block diagram of 3*3 Toffoli gate and (b) Equivalent quantum representation.

*D. Frekdin Gate*

The input vector, $I_v$ and output vector, $O_v$ for 3*3 Fredkin gate (FRG) [12] can be defined as follows: $I_v = (A, B, C)$ and $O_v = (P = A, Q = \overline{A}B \oplus AC, R = \overline{A}C \oplus AB)$. The quantum cost of Frekdin gate is 5. The block diagram and equivalent quantum representation for 3*3 Fredkin gate are shown in Fig. 4.

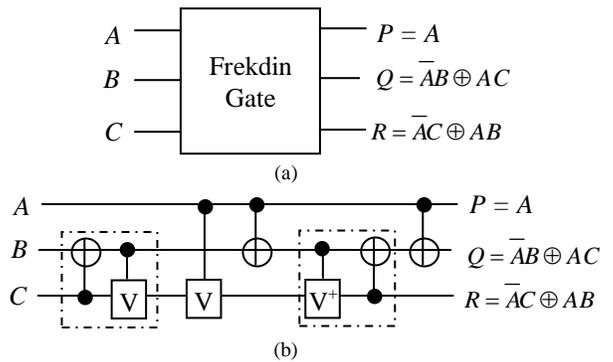

Fig. 4. (a) Block diagram of 3*3 Frekdin gate and (b) Equivalent quantum representation

*E. Peres Gate*

The input vector, $I_v$ and output vector, $O_v$ for 3*3 Peres gate (PG)[13] can be defined as follows: $I_v = (A, B, C)$ and $O_v = (P = A, Q = A \oplus B, R = AB \oplus C)$. The quantum cost of Peres gate is 4. The block diagram and equivalent quantum representation for 3*3 Peres gate are shown in Fig. 5.

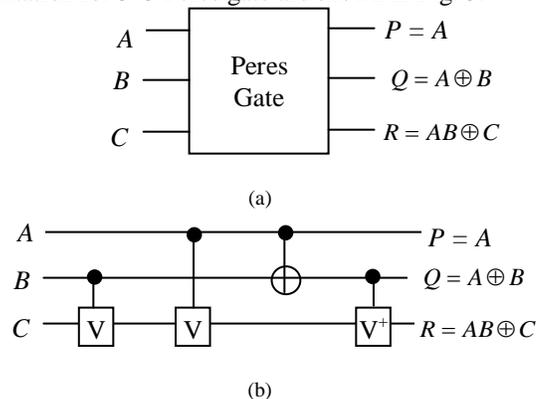

Fig. 5. Block diagram of 3*3 Peres and (b) Equivalent quantum representation

## IV. PROPOSED SAM GATE

After The input vector, $I_v$ and output vector, $O_v$ for 3*3 SAM Gate is defined as follows: $I_v$ = (A, B, C) and $O_v$ = ( $P = \bar{A}$, $Q = \bar{A}B \oplus A\bar{C}$, $R = \bar{A}C \oplus AB$ ). The block diagram of a 3*3 SAM gate is shown in Fig. 6.

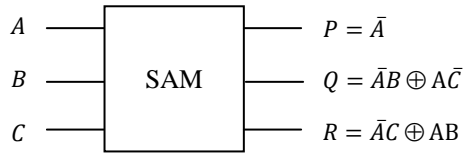

Fig. 6. Block diagram of a 3*3 SAM gate

The truth table for a 3x3 SAM gate is shown in Table I.

TABLE I. TRUTH TABLE FOR 3*3 SAM GATE

| A | B | C | $P = \bar{A}$ | $Q = \bar{A}B \oplus A\bar{C}$ | $R = \bar{A}C \oplus AB$ |
|---|---|---|---|---|---|
| 0 | 0 | 0 | 1 | 0 | 0 |
| 0 | 0 | 1 | 1 | 0 | 1 |
| 0 | 1 | 0 | 1 | 1 | 0 |
| 0 | 1 | 1 | 1 | 1 | 1 |
| 1 | 0 | 0 | 0 | 1 | 0 |
| 1 | 0 | 1 | 0 | 0 | 0 |
| 1 | 1 | 0 | 0 | 1 | 1 |
| 1 | 1 | 1 | 0 | 0 | 1 |

We can verify from the corresponding truth table of the SAM gate that the output and input vectors have one to one mapping between them which satisfies the condition of reversibility of a gate. We can see from Table I that the 8 different input and output vectors unique means they have one to one mapping them. So the proposed gate satisfies the condition of reversibility.

The Equivalent quantum representation of the SAM gate and minimization of quantum cost are shown in Fig 7(a) through 7(d). The quantum cost of SAM gate is 4.

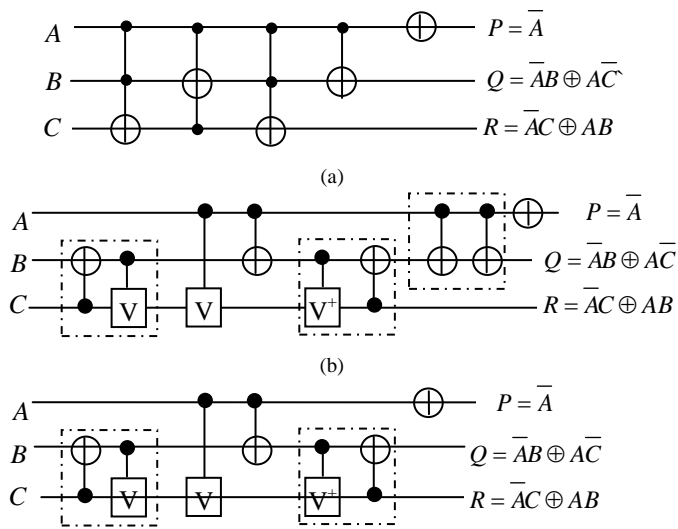

Fig. 7. Quantum cost of proposed SAM gate.

If we give 0 to 3rd input then we get NOT of 1st input in 1st output, OR or 1st and 2nd inputs in 2nd output and AND of 1st and 2nd inputs in 3rd output. This operation is shown in Fig. 8. So this gate can be used as two input universal gate.

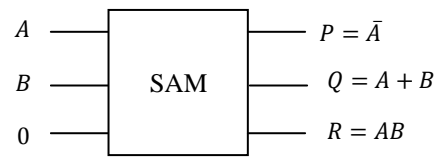

Fig. 8. SAM gate as two input universal gate.

## V. DESIGN AND SYNTHESIS OF REVERSIBLE SEQUENTIAL CIRCUITS

In this section, we presented novel designs of reversible flip-flops that are optimized in terms of quantum cost, delay and garbage outputs.

### A. The SR Flip-Flop

For SR flip-flop we modified the Peres gate. The modified Peres gate is shown in figure 9.

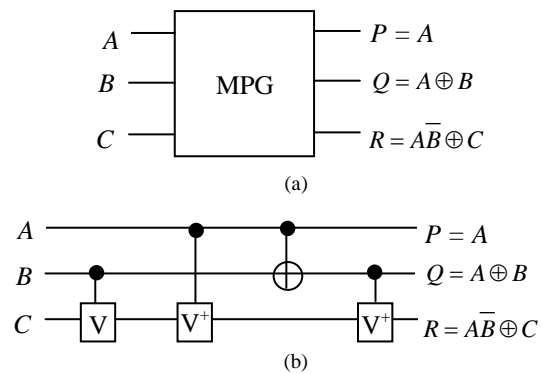

Fig. 9. (a) Block diagram of 3*3 MPG and (b) Equivalent quantum represenation

The characteristic equation of SR flip-flop is $Q = S + \overline{R}Q$. The SR flip-flop can be realized by a modified Peres gate (MPG). It can be mapped with the MPG by giving Q, R and S respectively in 1st, 2nd and 3rd inputs of the MPG. Fig. 10 shows the proposed design of SR flip-flop with $Q$ and $\overline{Q}$ outputs.

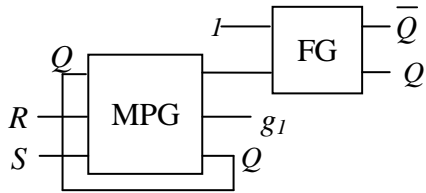

Fig. 10. Proposed design of SR flip-flop with $Q$ and $\overline{Q}$ outputs

The proposed SR flip-flop with $Q$ and $\overline{Q}$ outputs has quantum cost 5, delay 5 and has the bare minimum of 1 garbage bit. The proposed design of SR flip-flop achieves improvement ratios of 50% in terms of quantum cost, delay and garbage outputs compared to the design presented by Rice 2008 [14]. The improvement ratios compared to the design presented in Thapliyal et al.2010 [15] are 37%, 37% and 50% in terms quantum cost, delay and garbage outputs. The comparisons of our SR flip-flop (with $Q$ and $\overline{Q}$ outputs) design with existing designs in literature are summarized in Table II.

TABLE II. COMPARISONS OF DIFFERENT TYPES OF SR FLIP-FLOPS WITH Q AND $\overline{Q}$ OUTPUTS

| SR flip-flop design | Cost Comparison | | |
|---|---|---|---|
| | Quantum Cost | Delay | Garbage Outputs |
| Proposed | 5 | 5 | 1 |
| Existing [14] | 10 | 10 | 2 |
| Existing [15] | 8 | 8 | 2 |
| Improvement(%) w.r.t. [14] | 50 | 50 | 50 |
| Improvement(%) w.r.t. [15] | 37 | 37 | 50 |

This SR flip-flop design does not have enable signal (clock) and hence is not gated in nature. We proposed a design of gated SR flip-flop that can be realized by one MPG gate, one SAM and one FG gate. Another FG is needed to copy and produce the complement of Q. So we used a DFG instead of two FGs. The proposed gated SR flip-flop is shown in Fig. 11.

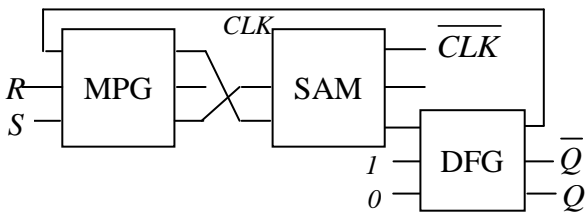

Fig. 11. Proposed design of gated SR flip-flop with $Q$ and $\overline{Q}$ outputs

Proposed gated SR flip-flop with $Q$ and $\overline{Q}$ outputs has quantum cost 10, delay 10 and has 2 garbage bits. The proposed design of gated SR flip-flop achieves improvement ratios of 41%, 41% and 33% in terms of quantum cost, delay and garbage outputs compared to the design presented in Thapliyal et al.2010 [15]. The comparisons of our gated SR flip-flop (with $Q$ and $\overline{Q}$ outputs) design with existing designs in literature are summarized in Table III.

TABLE III. COMPARISONS OF DIFFERENT TYPES OF GATED SR FLIP-FLOPS WITH Q AND $\overline{Q}$ OUTPUTS

| Gated SR flip-flop design | Cost Comparison | | |
|---|---|---|---|
| | Quantum Cost | Delay | Garbage Outputs |
| Proposed | 11 | 11 | 2 |
| Existing[15] | 17 | 17 | 3 |
| Improvement in (%) w.r.t. [15] | 41 | 41 | 33 |

Our proposed Master Slave SR flip-flop with only $Q$ output is shown in Fig. 12.

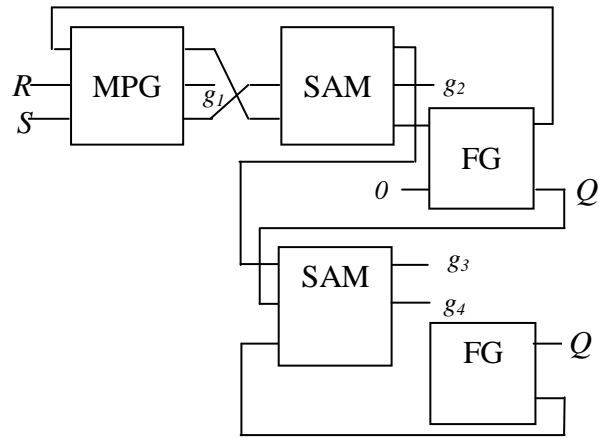

Fig. 12. Proposed Design of Master Slave SR flip-flop with only $Q$ output

The proposed master-slave SR flip-flop with only $Q$ output has quantum cost 14, delay 14 and has 4 garbage bits. The proposed design of master slave SR flip-flop achieves improvement ratios of 36% and 36% in terms of quantum cost and delay compared to the design presented in Thapliyal et al. 2010[15]. The comparisons of our Master Slave SR flip-flop design with existing designs in literature are summarized in Table IV.

TABLE IV. COMPARISONS OF DIFFERENT TYPES OF MASTER SLAVE SR FLIP-FLOPS WITH ONLY $Q$ OUTPUT

| Master slave SR flip-flop design | Cost Comparison | | |
|---|---|---|---|
| | Quantum Cost | Delay | Garbage Outputs |
| Proposed | 15 | 15 | 4 |
| Existing[15] | 22 | 22 | 4 |
| Improvement in (%) w.r.t. [15] | 36 | 36 | 0 |

*B. The JK Flip-Flop*

The characteristic equation of a JK flip-flop is $Q = J\overline{Q} + Q\overline{K}$. The JK flip-flop is realized by one SAM gate. It can be mapped with the SAM gate by giving *Q, J* and *K* to 1st, 2nd and 3rd inputs to the SAM gate. The proposed JK flip-flop with $Q$ and $\bar{Q}$ outputs is shown in Fig.13.

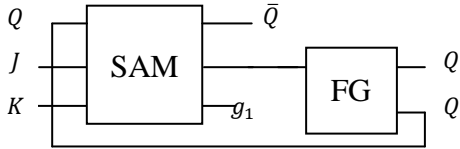

Fig. 13. Proposed design of JK flip-flop with $Q$ and $\bar{Q}$ outputs

The proposed JK flip-flop with $Q$ and $\bar{Q}$ outputs has quantum cost 5, delay 5 and has the bare minimum of 1 garbage bit. The proposed design of JK flip-flop achieves improvement ratios of 62%, 62% and 67% in terms of quantum cost, delay and garbage outputs compared to the design presented in Thapliyal et al. 2010[15]. The improvement ratios compared to the design presented in Lafifa Jamal et al. 2012[16] are 58%, 58% and 67% in terms quantum cost, delay and garbage outputs. The comparisons of our JK flip-flop (with Q and $\bar{Q}$ outputs) design with existing designs in literature are summarized in Table V.

TABLE V. COMPARISONS OF DIFFERENT TYPES OF JK FLIP-FLOPS WITH Q AND $\bar{Q}$ OUTPUTS

| JK flip-flop design | Cost Comparisons | | |
|---|---|---|---|
| | Quantum Cost | Delay | Garbage Outputs |
| Proposed | 5 | 5 | 1 |
| Existing[15] | 13 | 13 | 3 |
| Existing[16] | 12 | 12 | 3 |
| Improvement in (%) w.r.t. [15] | 62 | 62 | 67 |
| Improvement in (%) w.r.t. [16] | 58 | 58 | 67 |

The characteristic equation of gated JK flip-flop is $Q = \overline{CLK}Q + CLK(J\bar{Q} + Q\bar{K})$. The gated JK flip-flop with $Q$ and $\bar{Q}$ outputs is realized by two SAM gates and one DFG. The proposed gated JK flip-flop is shown in Fig.14.

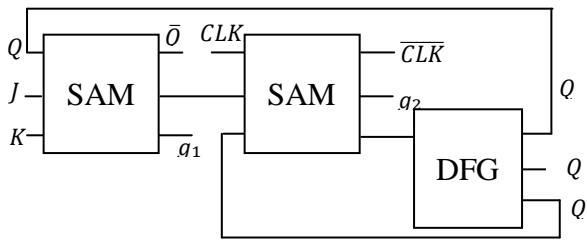

Fig. 14. Proposed Design of gated JK flip-flop with $Q$ and $\bar{Q}$ outputs.

The proposed gated JK flip-flop with $Q$ and $\bar{Q}$ outputs has quantum cost 10, delay 10 and has 2 garbage bits. The proposed design of gated JK flip-flop achieves improvement ratios of 37%, 37% and 33% in terms of quantum cost, delay and garbage outputs compared to the design presented in Thapliyal and Vinod 2007[17]. The improvement ratios compared to the design presented in Thapliyal et al. 2010[15] are 23%, 23% and 33% in terms quantum cost, delay and garbage outputs. The comparisons of our gated JK flip-flop (with $Q$ and $\bar{Q}$ outputs) design with existing designs in literature are summarized in Table VI.

TABLE VI. COMPARISONS OF DIFFERENT TYPES OF GATED JK FLIP-FLOPS WITH $Q$ AND $\bar{Q}$ OUTPUTS

| Gated JK flip-flop design | Cost Comparisons | | |
|---|---|---|---|
| | Quantum Cost | Delay | Garbage Outputs |
| Proposed | 10 | 10 | 2 |
| Existing[17] | 16 | 16 | 3 |
| Existing[15] | 13 | 13 | 3 |
| Improvement in (%) w.r.t. [17] | 37 | 37 | 33 |
| Improvement in (%) w.r.t. [15] | 23 | 23 | 33 |

Our proposed Master Slave JK flip-flop with $Q$ and $\bar{Q}$ outputs is shown in Fig.15.

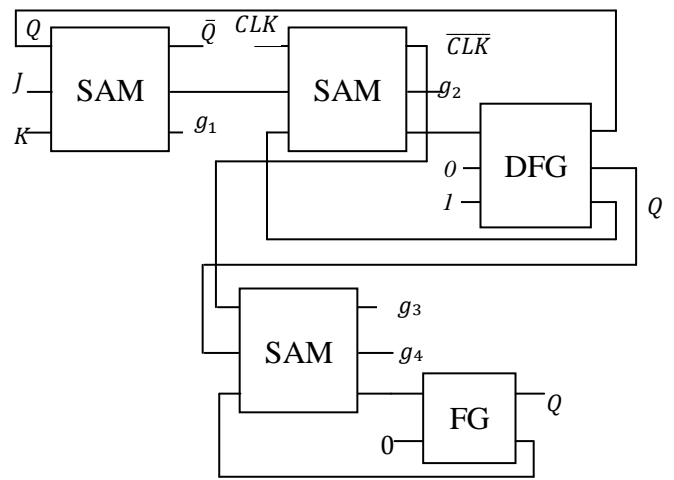

Fig. 15. Proposed Master Slave JK flip-flop with $Q$ and $\bar{Q}$ outputs

The proposed master-slave JK flip-flop with $Q$ and $\bar{Q}$ outputs has quantum cost 15, delay 15 and has 4 garbage bits. The proposed design of master slave JK flip-flop achieves improvement ratios of 37% and 37% in terms of quantum cost and delay compared to the design presented in Thapliyal and Vonod 2007[17]. The improvement ratios compared to the design presented in Thapliyal et al. 2010[15] are 21% and 21% in terms quantum cost and delay. The comparisons of our master slave JK flip-flop (with $Q$ and $\bar{Q}$ outputs) design with existing designs in literature are summarized in Table VII.

TABLE VII. COMPARISONS OF DIFFERENT TYPES OF MASTER SLAVE JK FLIP-FLOPS WITH $Q$ AND $\bar{Q}$ OUTPUTS

| Master slave JK flip-flop design | Cost Comparisons | | |
|---|---|---|---|
| | Quantum Cost | Delay | Garbage Outputs |
| Proposed | 15 | 15 | 4 |
| Existing[17] | 24 | 23 | 5 |
| Existing[15] | 19 | 19 | 4 |
| Improvement in (%) w.r.t. [17] | 37 | 37 | 20 |
| Improvement in (%) w.r.t. [15] | 21 | 21 | 0 |

*C. The D Flip-Flop*

The characteristic equation of gated D flip-flop is $Q = \overline{CLK}.Q + CLK.D$. The D flip-flop can be realized by one SAM gate and one DFG. It can be mapped with SAM gate by giving CLK, D and Q respectively in $1^{st}$, $2^{nd}$ and $3^{rd}$ inputs of SAM gate. The Fig. 16 shows our proposed gated D flip-flop with with $Q$ and $\overline{Q}$ outputs.

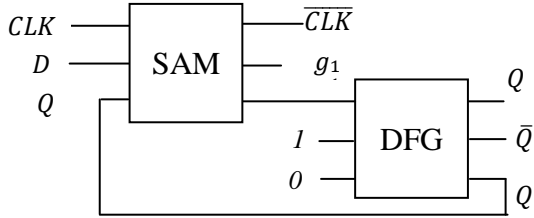

Fig. 16. Proposed design gated D flip-flop with $Q$ and $\overline{Q}$ outputs.

The proposed gated D flip-flop with $Q$ and $\overline{Q}$ outputs has quantum cost 6, delay 6 and has the bare minimum of 1 garbage bit. The proposed design of gated D flip-flop achieves improvement ratios of 14%, 14% and 50% in terms of quantum cost, delay and garbage outputs compared to the design presented in Thapliyal et al. 2010[15] and Lafifa Jamal et al. 2012[16]. The comparisons of our gated D flip-flop (with $Q$ and $\overline{Q}$ outputs) design with existing designs in literature are summarized in Table VIII.

TABLE VIII. COMPARISONS OF DIFFERENT TYPES OF GATED D FLIP-FLOPS WITH Q AND $\overline{Q}$ OUTPUTS

| D flip-flop design | Cost Comparisons | | |
|---|---|---|---|
| | *Quantum Cost* | *Delay* | *Garbage Outputs* |
| Proposed | 6 | 6 | 1 |
| Existing[15] | 7 | 7 | 2 |
| Existing[16] | 7 | 7 | 2 |
| Improvement in (%) w.r.t. [15] | 14 | 14 | 50 |
| Improvement in (%) w.r.t. [16] | 14 | 14 | 50 |

Our proposed Master Slave D flip-flop with $Q$ and $\overline{Q}$ outputs is shown in Fig. 17.

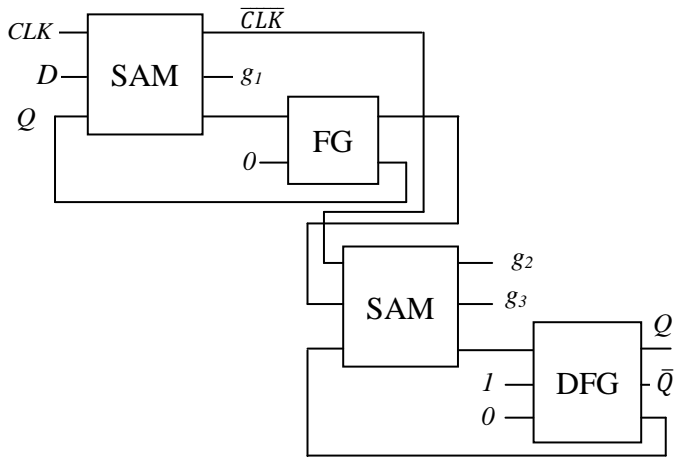

Fig. 17. Proposed design Master Slave D flip-flop with $Q$ and $\overline{Q}$ outputs.

The proposed master-slave D flip-flop with $Q$ and $\overline{Q}$ outputs has quantum cost 11, delay 11 and has 3 garbage bits. The proposed design of master slave D flip-flop achieves improvement ratios of 21% and 21% in terms of quantum cost and delay compared to the design presented in Chuang et al. 2008[18]. The improvement ratios compared to the design presented in Thapliyal et al. 2010[15] is 21% and 21% in terms quantum cost and delay. The comparisons of our master slave D flip-flop (with $Q$ and $\overline{Q}$ outputs) design with existing designs in literature are summarized in Table IX.

TABLE IX. COMPARISONS OF DIFFERENT TYPES OF MASTER SLAVE D FLIP-FLOPS WITH $Q$ AND $\overline{Q}$ OUTPUTS

| Master Slave D flip-flop design | Cost Comparisons | | |
|---|---|---|---|
| | *Quantum Cost* | *Delay* | *Garbage Outputs* |
| Proposed | 11 | 11 | 3 |
| Existing[18] | 14 | 14 | 3 |
| Existing[15] | 13 | 13 | 3 |
| Improvement in (%) w.r.t. [17] | 21 | 21 | 0 |
| Improvement in (%) w.r.t. [15] | 15 | 15 | 0 |

VI. CONCLUSION

Reversible latches are going to be the main memory block for the forthcoming quantum devices. In this paper we proposed optimized reversible D latch and JK latches with the help of proposed SAM gates. Appropriate algorithms and theorems are presented to clarify the proposed design and to establish its efficiency. We compared our design with existing ones in literature which claims our success in terms of number of gates, number of garbage outputs and delay. This optimization can contribute significantly in reversible logic community.

ACKNOWLEDGMENT

The authors would like to thank the anonymous referees for their supports and constructive feedback, which helped significantly to improve technical quality of this paper.